\documentclass[aps,prb,reprint,groupedaddress,showpacs]{revtex4-1}
\usepackage{amsfonts}
\usepackage{amsmath}
\usepackage{amssymb}
\usepackage{graphicx}%
\usepackage{graphicx}% Include figure files
\usepackage{epsfig}
\begin{document}

%\preprint{}

\title{First-passage theory of exciton population loss in single-walled carbon nanotubes reveals micron-scale intrinsic diffusion lengths}

\author{Mitchell D. Anderson}
\affiliation{Department of Physics, Engineering Physics $\&$ Astronomy, Queen's University, Kingston, Ontario, K7L 3N6 Canada }
\author{Yee-fang Xiao}
\affiliation{Department of Physics, Engineering Physics $\&$ Astronomy, Queen's University, Kingston, Ontario, K7L 3N6 Canada }
\author{James M. Fraser}
\email{fraser@physics.queensu.ca}
\affiliation{Department of Physics, Engineering Physics $\&$ Astronomy, Queen's University, Kingston, Ontario, K7L 3N6 Canada }%

\date{\today}

\begin{abstract}
One-dimensional crystals have long range translational invariance which manifests as long exciton diffusion lengths, but such intrinsic properties are often obscured by environmental perturbations.
We use a first-passage approach to model single-walled carbon nanotube (SWCNT) exciton dynamics (including exciton-exciton annihilation and end effects) and compare it to results from both continuous-wave and multi-pulse ultrafast excitation experiments to extract intrinsic SWCNT properties.  
Excitons in suspended SWCNTs experience macroscopic diffusion lengths, on the order of the SWCNT length,  (1.3-4.7 $\mu$m) in sharp contrast to encapsulated samples. 
For these pristine samples, our model reveals intrinsic lifetimes (350-750 ps), diffusion constants (130-350 cm$^{2}$/s), and absorption cross-sections (2.1-3.6 $\times 10^{-17}$ cm$^{2}$/atom) among the highest previously reported.
\end{abstract}

%\keywords{}

\maketitle

%\section{}
%\subsection{}
%\subsubsection{}

%%%%%%%%%%%%%%%%%%%%%%%%%%%%%%%%%%%%%%%%%%%%%%%%%%%%%%%%INTRO%%%%%%%%%%%%%%%%%%%%%%%%%%%%%%%%%%%%%%%%%%%%%%%%%%%%%%%%%%%%%%%%%%%%%%%%%%%%%%%%%%%%%%%%%%%%%%%%%%%%%%

Semiconductor nanoscience has rapidly expanded since material engineering enabled control of the dimensionality, with two dimensional (quantum well), one dimensional (quantum wire), and zero dimensional (quantum dot) systems now readily producible. 
These nanosystems have been particularly useful for optical applications due to their customizable electronic properties and, in contrast to bulk semiconductors, the electronic density of states (DOS) at the bandgap is non-zero.
In the case of quantum wires, the one-dimensional quantum confinement leads to Van Hove singularities, which are divergences in the DOS at the bandgap.
An example of a nearly ideal quantum wire is the single-walled carbon nanotube (SWCNT), which is a rolled graphene sheet with radial quantum confinement and translation invariance along the SWCNT axis. 
The one-dimensionality of carriers in SWCNTs results in a strong Coulomb interaction which binds the electron-hole pair into an exciton with a binding energy approximately one third of the single particle bandgap \cite{Perebeinos_PRL04,Wang_Sci05}.

For many SWCNT samples, the exciton lifetime is dominated by end quenching, defects/adsorbates, and, when highly excited, interactions between excitons. 
These decay mechanisms are all strongly dependent on exciton transport, which is often modeled as diffusive, with an effective diffusion length determined by the sample quality\cite{Cognet_Sci07,Liu_JPhysChemC11}.
The high likelihood of spatial overlap between diffusing excitons results in efficient Auger-like exciton-exciton annihilation (EEA) which limits the maximum exciton density well below the predictions of phase-space filling\cite{Ma_PRL05}.
In addition to EEA, diffusion also allows quenching of excitons at the ends and defect sites, severely limiting the quantum efficiency for many samples.  
Interestingly, both diffusion driven end/defect quenching and EEA enable many realized and potential photonic applications \cite{Huang_PRL06}. 
At low exciton densities, the stochastic fluctuations of the optical emission intensity due to the adsorption and desorption of fluorescence quenchers has led to single molecule counting arrays and biological sensors \cite{Barone_NatMat04,Kim_NatureChem09,Ulissi_JPhysChemLett11}. 
At high exciton densities, the rapid relaxation of excitons due to EEA may give rise to photon anti-bunching, showing promise for single-photon sources with selectable wavelengths from optical into the telecom band \cite{Hogele_PRL08}. 

Since a SWCNT is all surface, it is particularly sensitive to environmental perturbations, such as adsorbates.
For this reason, the underlying one-dimensional translation symmetry of the nanostructure can be obscured and many of the intrinsic SWCNT properties are still elusive. 
For this reason, results from recent experimental studies vary by orders of magnitude, for example optical absorptions (0.02 - 2.5 $\times$ 10$^{-17}$ cm$^{2}$/atom)\cite{Harrah_Nano11,Joh_NatNano11,Schoppler_JPhysChemC11}, lifetimes (3 - 350 ps)\cite{Berciaud_PRL08,Ruzicka_PRB12}, and diffusion constants (0.1-300 cm$^{2}$/s)\cite{Harrah_Nano10,Ruzicka_PRB12}.
Such results provide little constraint on the exciton diffusion range, placing it somewhere between the carbon inter-atomic spacing to the typical SWCNT length.  
The correct estimation of an ``upper bound" for the diffusion length is interesting as both a fundamental science question as well as from an applications point of view. 
If intrinsic diffusion lengths correspond to SWCNT length, exciton dynamics would be strongly affected by nonlinear effects (EEA) even at low exciton numbers. 
Experimental results for suspended SWCNTs show exciton interactions at very low densities, making current (thermodynamic limit) diffusion models invalid and calling into question our basic understanding of exciton dynamics.\cite{Murakami_PRL09,Xiao_PRL10,Moritsubo_PRL10,Xie_PRB12} 

%%%%%%%%%%%%%%%%%%%%%%%%%%%%%%%%%%%%%%%%%%%%%%%%%%%%%%%%%%%%%%%%%%%%%%%%%%%%%%%%%%%%%%%%%%%%%%%%%%%%%%%%%%%%%%%%%%%%%%%%%%%%%%%%%%%%%%%%%%%%%%%%%%%%%%%%%%%%%%%%%%%%%%%%%%

%%% NEW
While there has been much work on the diffusion of excitons in SWCNTs, it has relied primarily on solutions to the standard differential diffusion equation \cite{Rajan_PhysChemB08,Hertel_Nano10,Siitonen_NL10,Siitonen_NL12}. 
This approach is ``valid only if we ignore effects which happen in time intervals of the order of [the scattering time] and space intervals of the order of [the mean free path]"\cite{Naqvi_PRL82}, and begins to break down for pristine SWCNTs, when the scattering time becomes long and the diffusivity becomes high. In addition, steady state solutions are often used, which are inappropriate for use in pulsed excitation experiments. An alternate approach to calculate population loss is first-passage theory including an absorbing boundary\cite{Harris_JPhysA80,Harris_JCHEMPHYS80,Harris_PRA87,Bray_PRL02}, which we will show, has the advantage of easily incorporating EEA even at low densities.

%%%
In this article we develop an analytic model using a first-passage approach of the exciton survival probabilities, including end quenching and EEA. 
An important feature of our model, is the finite line segment pairwise distribution function (instead of the ring pairwise distribution function used in previous work), making the nonlinear interactions valid below the thermodynamic limit. 
We use the analytic model to extract intrinsic SWCNT properties from both continuous-wave (CW) and pulsed ultrafast photoluminescence (PL) experiments on individual suspended SWCNTs. 
Monte Carlo simulations are used to verify the model for typical SWCNT parameters and a range of injected exciton densities.  
We find strong evidence to support the assertion that unprocessed suspended SWCNTs are the most ``pristine'' of any sample; they exhibit long intrinsic diffusion lengths, indicative of the expected long range translational invariance of the underlying structure.
We extract exciton diffusion lengths of several microns, on the order of the SWCNT length, which arises from the long intrinsic lifetimes (350-750 ps) and high mobilities (130 - 350 cm$^{2}$/s) of these samples.
We also show that suspended SWCNTs exhibit the highest optical absorptions (2.1-3.6 $\times 10^{-17}$ cm$^{2}$/atom) of any sample, approximately 5 times larger than graphene.  

%%%%%%%%%%%%%%%%%%%%%%%%%%%%%%%%%%%%%%%%%%%%%%%%%%%%%%%%%%%%%%%%%%%%%%%%%%%%%%%%%%%%%%%%%%%%%%%%%%%%%%%%%%%%%%%%%%%%%%%%%%%%%%%%%%%%%%%%%%%%%%%%%%%%%%%%%%%%%%%%%%%%%%%%%%%%%%%%%%%%%%%%%%%

A key input to this analysis are experimental results from two different types of studies: PL dependence on SWCNT length in the low fluence regime and PL dependence on pulse delay for ultrafast excitation ($\sim$ 150 fs).
For the former, model fits to the population loss to the SWCNT ends (as a function of suspended SWCNT length) allows extraction of the intrinsic exciton diffusion length. 
In contrast, application of the model to ultrafast double-pulse excitation (i.e., femtosecond excitation correlation spectroscopy (FEC)\cite{MiyauchiFEC_PRB09,Xiao_PRL10}) allows extraction of the intrinsic lifetime ($\tau$) and the diffusivity ($D_{X}$), which is used to calculate the intrinsic diffusion length $L_{D}=\sqrt{D_{X}\tau}$. Using the extracted exciton diffusion lengths, we are able to use PL saturation from single pulse excitation experiments to determine the optical absorption cross-section ($\sigma_{ab}$).
All experimental studies use similar samples: single suspended SWCNTs grown by chemical vapor deposition on lithographically patterned silicon dioxide on silicon wafers. During experimentation, samples are in air, dry N$_{2}$, or in vacuum ($\sim 10^{-6}$ mbar).  PL excitation spectroscopy and high resolution PL mapping of candidate SWCNT allows determination of species, quality, length, and orientation to ensure study of pristine SWCNTs. 

%%%%%%%%%%%%%%%%%%%%%%%%%%%%%%%%%%%%%%%%%%%%%%%%%%%%%%%%%%%%%%%%%%%%%%%%%%%%%%%%%%%%%%%%%%%%%%%%%%%%%%%%%%%%%%%%%%%%%%%%%%%%%%%%%%%%%%%%%%%%%%%%%%%%%%%%%%%%%%%%%%%%%%%%%%%%%%%%%%%%%%%%%%%

The analytic model determines the time-dependent exciton survival probability including the three main decay processes, linear decay, decay due to EEA, and decay due to end quenching, as characterized by their respective survival probabilities: $P_{\Gamma}(t)$, $P_{\rm{EEA}}(N_{0},t)$ and $P_{\rm{END}}(t)$, where $N_{\rm{0}}$ is the initial exciton number. 
Calculation of the total survival probability, $P_{\rm{TOT}}(N_{0},t)=P_{\Gamma}(t) \times P_{\rm{EEA}}(N_{0},t) \times P_{\rm{END}}(t)$, allows the quantum efficiency to be determined from $\eta_{\rm{QE}} = \Gamma_{\rm{R}}\int_{0}^{\infty}{P_{\rm{TOT}}(N_{0},t)\rm{dt}}$, where $\Gamma_{\rm{R}}$ is the radiative decay rate\cite{MiyauchiFEC_PRB09}. Monte Carlo numerical simulations are used to verify this model.

%%%%%%%%%%%%%%%%%%%%%%%%%%%%%%%%%%%%%%%%%%%%%%%%%%%%%%%%%%%%%%%%%%%%%%%%%%%%%%%%%%%%%%%%%%%%%%%%%%%%%%%%%%%%%%%%%%%%%%%%%%%%%%%%%%%%%%%%%%%%%%%%%%%%%%%%%%%%%%%%%%%%%%%%%%%%%%%%%%%%%%%%%%%

It is useful to identify the intrinsic quantum efficiency of a SWCNT, i.e., total photons emitted divided by photons absorbed in the regime where end effects and EEA are negligible ($P_{\rm{END}}\times P_{\rm{EEA}} \rightarrow 1$) and survival probability is dominated by linear decay: $P_{\Gamma}(t)=e^{-\Gamma t}$, with $\Gamma=\Gamma_{R}+\Gamma_{NR}$, where $\Gamma_{R}$ and $\Gamma_{NR}$ are radiative and non-radiative decay respectively. For this paper the intrinsic lifetime is defined from this relationship with $\tau=1/\Gamma$.
The intrinsic non-radiative decay is predicted to be dominated by multiphonon decay and phonon-assisted indirect exciton ionization \cite{Perebeinos_PRL08}.
In this limit, exciton population follows a standard mono-exponential \cite{Hagen_PRL05}, $P_{\rm{TOT}}(t)=e^{-\Gamma t}$, so intrinsic quantum efficiency can be written as $\eta_{0} = \Gamma_{R}/\Gamma$.

%%%%%%%%%%%%%%%%%%%%%%%%%%%%%%%%%%%%%%%%%%%%%%%%%%%%%%%%%%%%%%%%%%%%%%%%%%%%%%%%%%%%%%%%%%%%%%%%%%%%%%%%%%%%%%%%%%%%%%%%%%%%%%%%%%%%%%%%%%%%%%%%%%%%%%%%%%%%%%%%%%%%%%%%%%%%%%%%%%%%%%%%%%%

Experimental results showing that the PL action cross-section ($\sigma_{\rm{ab}}\times \eta_{\rm{QE}}$) strongly depends on the SWCNT length imply that end quenching must be explicitly included as a decay channel even in micron long SWCNTs under tightly focused excitation conditions \cite{Moritsubo_PRL10}.
The diffusion-quenching process can be modeled using the first passage time of a Brownian motion\cite{Karatzas_BMSS91}. 
In this usage the probability that a diffusive particle has not been quenched by a site, at an initial distance $x$, after time $t$ is $P(x,t)=\rm{erf}(x/2\sqrt{D_{X}t})$.
Therefore, the probability that an exciton, generated at a random location $x$ on a SWCNT of length $L$, survives end quenching is $P(L/2+x,t)\times P(L/2-x,t)$.
The injected position probability scales with the intensity profile so that the mean probability of an exciton surviving end quenching is
\begin{equation}		
	P_{\rm{END}}=\frac{\int\limits_{-L/2}^{L/2}g(x)P(L/2+x,t)P(L/2-x,t)dx}{\int\limits_{-L/2}^{L/2}{g(x)dx}}
	\label{eq:three}
\end{equation}
where $g(x)$ is the spatial weighting function of exciton injection determined by the intensity profile of the excitation source. 
In the case of homogeneous excitation, $P_{\rm{END}}$ can be solved analytically, 

\begin{equation}	
\begin{split}
P_{\rm{END}}(\tau_{D},t)=&\left[\mathrm{erf}\left(\frac{1}{2}\sqrt{\frac{\tau_{D}}{2t}}\right)\right]^{2}-\sqrt{\frac{32t}{\pi\tau_{D}}} \times \\& \left[\frac{1}{\sqrt{2}}\mathrm{erf}\left(\sqrt{\frac{\tau_{D}}{4t}}\right)-\mathrm{erf}\left(\sqrt{\frac{\tau_{D}}{8t}}\right)e^{-\frac{\tau_{D}}{8t}}\right]
\end{split}
	\label{eq:edgediffusion}
\end{equation}

revealing that the decay dynamics for a given SWCNT depends only on the ``diffusional time'', $\tau_{D}=L^{2}/D_{X}$. While this solution is for two mobile quenching sites\cite{Bray_PRL02}, it works better in the parameter space of interest than the solution for fixed quenching sites which is accurate for $t >> \tau_{D}$.
Alternatively, for comparison with SWCNT vs length studies, instead of the diffusional time a unitless parameter $\zeta = L/L_{D}$ and the intrinsic lifetime, $\tau$, can be used.
This formulation is useful not only in understanding the PL vs SWCNT length trends but, as we will show, also accurately predicts population decay dynamics without the need of a phenomenological bi-exponential model \cite{Miyauchi_JPhysChemC10}.

%%%%%%%%%%%%%%%%%%%%%%%%%%%%%%%%%%%%%%%%%%%%%%%%%%%%%%%%%%%%%%%%%%%%%%%%%%%%%%%%%%%%%%%%%%%%%%%%%%%%%%%%%%%%%%%%%%%%%%%%%%%%%%%%%%%%%%%%%%%%%%%%%%%%%%%%%%%%%%%%%%%%%%%%%%%%%%%%%%%%%%%%%%%

\begin{figure}
  \includegraphics[width=0.4\textwidth]{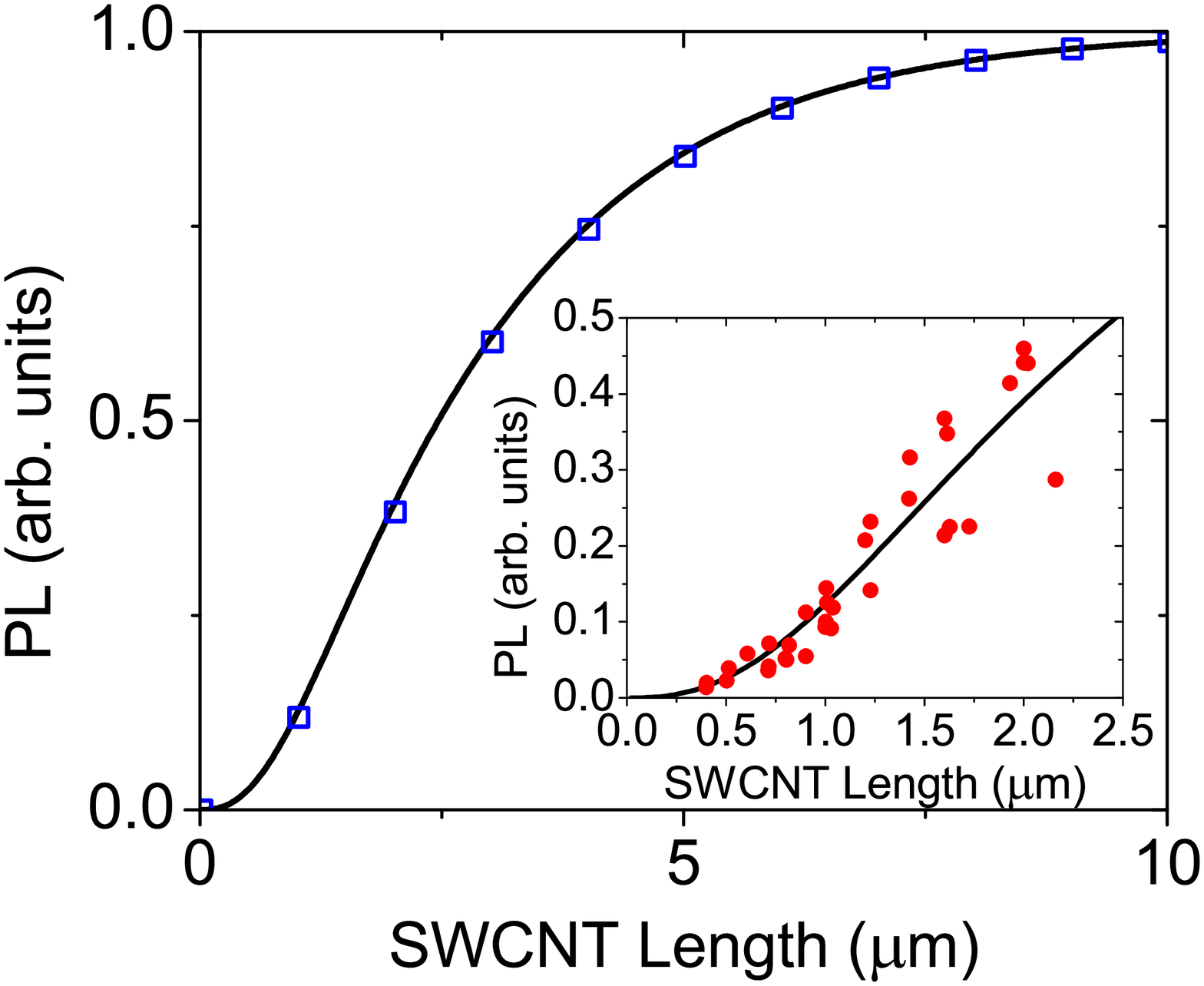} 
  \caption{Model prediction of PL dependence on SWCNT length in the linear regime ($P_{\rm{EEA}}=1$) for excitons with a 1.3 $\mu$m diffusion length, excited with a Gaussian beam ($\sigma$ = 520 nm), blue squares show Monte Carlo simulations. [INSET] Detail of model results with experimental data (red circles) collected by Moritsubo \textit{et al.}\cite{Moritsubo_PRL10}.}
  \label{fig:PLvsL}
\end{figure}

%%%%%%%%%%%%%%%%%%%%%%%%%%%%%%%%%%%%%%%%%%%%%%%%%%%%%%%%%%%%%%%%%%%%%%%%%%%%%%%%%%%%%%%%%%%%%%%%%%%%%%%%%%%%%%%%%%%%%%%%%%%%%%%%%%%%%%%%%%%%%%%%%%%%%%%%%%%%%%%%%%%%%%%%%%%%%%%%%%%%%%%%%%%

Previous work by \citet{Moritsubo_PRL10} showed increasing PL as a function of suspended segment length, excited with a centered Gaussian beam, $\sigma$ = 520 nm.  
To fit these experimental results, the quantum efficiency is calculated with $P_{\rm{TOT}}(t)=P_{\Gamma}(t) \times P_{\rm{END}}(t)$ (since with low intensity CW excitation, EEA is negligible). 
The exciton injection probability depends on the SWCNT length and excitation intensity profile:  $P_{\rm{inj}}=\int_{-L/2}^{L/2}g(x)dx/\int_{-\infty}^{\infty}g(x)dx$, so the total PL is $\propto P_{\rm{inj}}\times\eta_{QE}$. 
The experimental data is then fit with the \textit{intrinsic} diffusion length and a y-scaling parameter, the latter could be avoided if data included longer SWCNT segments. 
For the best fit (Figure \ref{fig:PLvsL} (INSET)), the extracted mean \textit{intrinsic} exciton diffusion length is $\sim$ 1.3 $\mu$m, significantly longer than the \textit{effective} diffusion length found previously \cite{Moritsubo_PRL10} but consistent with recent predictions \cite{Crochet_NL12}. 
The difference in the extracted diffusion length is believed to result from a limitation of the differential equation approach, which ``does not account for the spatial and $L_{D}$ dependence of $\tau$"\cite{Moritsubo_PRL10}. PL emission continues to increase even for SWCNTs much longer than $\sim$ 1.3 $\mu$m as shown in Figure \ref{fig:PLvsL}. 
End quenching is more effective than expected from a naive estimate based on diffusion length relative to SWCNT length, due to the high recurrence rates of diffusion in 1D systems.  

%%%%%%%%%%%%%%%%%%%%%%%%%%%%%%%%%%%%%%%%%%%%%%%%%%%%%%%%%%%%%%%%%%%%%%%%%%%%%% EEA %%%%%%%%%%%%%%%%%%%%%%%%%%%%%%%%%%%%%%%%%%%%%%%%%%%%%%%%%%%%%%%%%%%%%%%%%%%%%%%%%%%%%%%%%%%%%%
To model the saturation of PL under ultrafast excitation\cite{Xiao_PRL10}, we need to move beyond the linear regime and include exciton-exciton effects (i.e., EEA).
The relative diffusion constant of two diffusing particles is simply 2$D_{X}$. Therefore the probability that an exciton pair, initially separated by a distance $L_{X}$, has not undergone EEA is $\mathrm{erf}(L_{X}/2\sqrt{2D_{X}t})$.  
The EEA survival probability is thus highly dependent on the initial pair separation $L_{X}$ which is defined by the geometry. 
Here we use a finite line segment with the linear pair distribution function\cite{Tonks_PR36}, $f(L_{X},N_{0})= L^{N_{0}}\left(L-L_{X}\right)^{N_{0}-1}/N_{0}$, instead of the previously used ring geometry. 
The probability that a single exciton pair survives annihilation after time $t$ is then,
\begin{equation}
	P_{\rm{EEA}}^{'}=\int\limits_{0}^{L}f(L_{X},N_{0})\mathrm{erf}\left(\frac{L_{X}}{2\sqrt{2D_{X}t}}\right) dL_{X}
\end{equation}
This is appropriate for both high and low exciton densities, in contrast to a ring geometry which is only valid in the thermodynamic limit \cite{Srivastava_PRB09}.
To find the total exciton population given $N_{0}$ injected excitons we write, $N(t)=1+(N_{0}-1)P_{\rm{EEA}}^{'}$. 
The total probability an exciton will survive is $N(t)/N_{0}$, which has the solution,
\begin{equation}
\begin{split}
	 P_{\rm{EEA}} = &\frac{1}{N_{0}}\left[1+\frac{N_{0}-1}{N_{0}+1} \sqrt{\frac{\tau_{D}}{2\pi t}} \times \right.  \\ &
	 \left. F\left(\left[\frac{1}{2},1 \right],\left[1+\frac{N_{0}}{2},\frac{3}{2}+\frac{N_{0}}{2} \right], -\frac{\tau_{D}}{8 t} \right) \right]
\end{split}
\end{equation}
where F(n,d,z) is the generalized hypergeometric function.  This shows a very convenient result: similar to the linear regime, the survival probability only depends on $\tau_{D}$.  

%%%%%%%%%%%%%%%%%%%%%%%%%%%%%%%%%%%%%%%%%%%%%%%%%%%%%%%%%%%%%%%%%%%%%%%%%%%%%%%%%%%%%%%%%%%%%%%%%%%%%%%%%%%%%%%%%%%%%%%%%%%%%%%%%%%%%%%%%%%%%%%%%%%%%%%%%%%%%%%%%%%%%%%%%%%%%%%%%%%%%%%%%%%

\begin{figure}
	\includegraphics[width=0.4\textwidth]{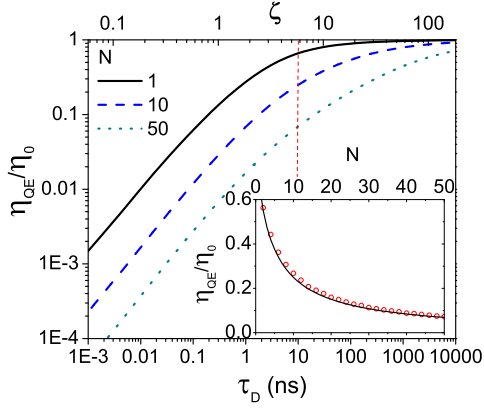}
  \caption{Effective SWCNT quantum efficiency for N injected excitons (using a linear pair distribution function), for SWCNTs over a range of diffusional times, $\tau_{D}=L^{2}/D_{X}$ and the unitless parameter, $\zeta=\sqrt{\tau_{D}/\tau}=L/L_{D}$ with $\tau =$ 300 ps. [INSET] Effective quantum efficiency for increasing injected exciton number($\tau_{D}$ = 10 ns, corresponding to red dashed line in main figure) compared to Monte Carlo simulation results(red circles).}
  \label{fig:edgediffusion}
\end{figure}

Total survival probability was assumed to be $P_{\rm{TOT}}=P_{\Gamma}\times P_{\rm{EEA}}\times P_{\rm{END}}$, which is exactly correct only when the survival probabilities from the different decay channels are independent.
For typical SWCNT lengths and diffusivities, this approximation is verified by comparison by Monte Carlo simulations: PL and population dynamics agree to a few percent over the parameter range of interest(\textit{e.g.} Figure \ref{fig:edgediffusion} INSET).

As observed experimentally\cite{Moritsubo_PRL10}, we show that for typical SWCNTs ($\tau_{D}$= 0.1 - 10 ns) the effective quantum efficiency can be drastically lower than the intrinsic quantum efficiency due to end quenching. 
Additionally, the linear one dimensional geometry of the SWCNT results in the quantum efficiency being significantly reduced by EEA, even for low exciton injection numbers as shown in Figure \ref{fig:edgediffusion}.  
Previous work that assumed a ring geometry underestimated this effect for low exciton numbers since excitons are more likely to be injected closer to each other on a line segment than on a ring.

%%%%%%%%%%%%%%%%%%%%%%%%%%%%%%%%%%%%%%%%%%%%%%%%%%%%%%%%%%%%%%%%%%%%%%%%%%%%%%%%%%%%%%%%%%%%%%%%%%%%%%%%%%%%%%%%%%%%%%%%%%%%%%%%%%%%%%%%%%%%%%%%%%%%%%%%%%%%%%%%%%%%%%%%%%%%%%%%%%%%%%%%%%%

\begin{figure}
  \includegraphics[width=0.48\textwidth]{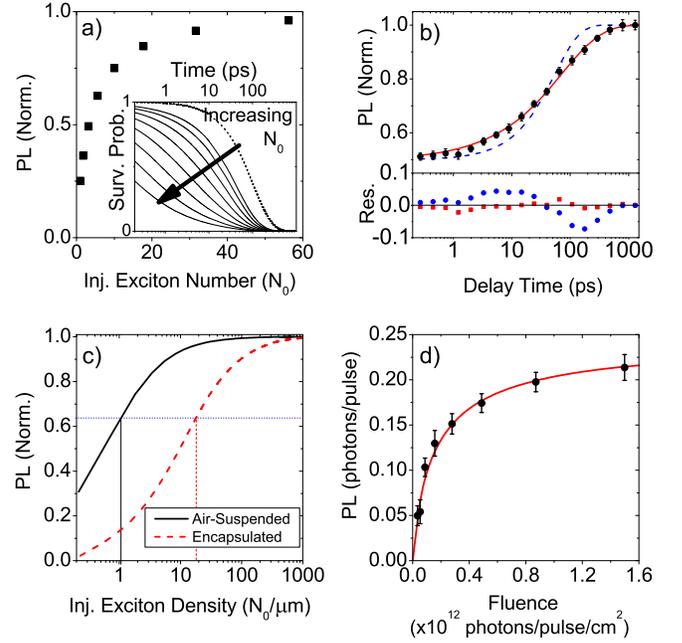}
  \caption{(a) PL for increasing initial exciton number with [INSET] corresponding exciton population dynamics. Each exciton survival probability curve in the inset corresponds to a point on the main figure. Purely linear decay (dashed line) noted for comparison. (b) FEC data (black circles)\cite{Xiao_PRL10} fit with mono-exponential decay (blue dashed curve) and diffusion model (red solid curve), and (bottom) corresponding residues. (c) Calculated PL for a typical encapsulated (red dashed curve) and air-suspended SWCNT (black solid curve) with saturation points as defined in text. (d) Experimentally measured PL for a 4.6 $\mu$m (9,8) air-suspended SWCNT (black circles) and best fit (red curve).}
  \label{fig:PPD}
\end{figure}

% use the term FEC at the beginning here.
When the injected exciton number is high, the EEA pathway dominates and a saturation of the PL occurs, as seen in Figure \ref{fig:PPD} a).
In FEC, this saturation is exploited to extract the intrinsic lifetimes.  Two pump pulses, separated by a variable delay time ($\tau_{x}$), each excite the SWCNT into the saturation regime.  
The longer the delay between pulses, the greater the total PL emitted.  This characteristic FEC signal can be fit using,
\begin{equation}
	PL_{\rm{FEC}}(\tau_{x})=\frac{\int\limits_{0}^{\tau_{x}} P(t)\rm{dt} + \int\limits_{0}^{\infty} P(t)\rm{dt}}{2\int\limits_{0}^{\infty} P(t)\rm{dt}}
	\label{eq:FEC3}
\end{equation}
where $P(t)=P_{\rm{TOT}}$ contains the extractable parameters, $D_{X}$ and $\Gamma$, and is independent of $N_{0}$ provided the pump is in the saturation regime.
An example is shown in Figure \ref{fig:PPD} (b), of a 4.6 $\mu$m (9,8) SWCNT, which has an intrinsic lifetime of 350 ps and a diffusivity of 180 cm$^{2}$/s, this very high diffusivity is consistent with recent observations \cite{Ruzicka_PRB12}.
It should be noted that this model has significantly better agreement than the standard mono-exponential decay (Figure \ref{fig:PPD} (b)) and does not require the additionally free parameters of a phenomenological bi-exponential decay model\cite{Miyauchi_JPhysChemC10}. 
Seven SWCNTs were studied and both the intrinsic lifetimes (350-750 ps) and diffusivities (130 - 350 cm$^{2}$/s) are among the highest of any SWCNT sample.
The corresponding diffusion lengths ($2.4-4.7 \mu$m) are on the same order as the CW PL vs length study in the linear regime done above, $\sim$ 1.3 $\mu$m.
This higher diffusion length for ultrafast experiments could be due to the higher energy of non-thermalized excitons undergoing EEA in the many exciton regime.
It should be noted that in both cases we are quantifying the \textit{intrinsic} diffusion length, which one would expect to observe on a pristine infinitely long SWCNT. 
Since the SWCNT length is often shorter than the intrinsic diffusion length, the effective diffusion length may be significantly shorter due to the shorter effective lifetime caused by end quenching.
%%%%%%%%%%%%%%%%%%%%%%%%%%%%%%%%%%%%%%%%%%%%%%%%%%%%%%%%%%%%%%%%%%%%%%%%%%%%%%%%%%%%%%%%%%%%%%%%%%%%%%%%%%%%%%%%%%%%%%%%%%%%%%%%%%%%%%%%%%%%%%%%%%%%%%%%%%%%%%%%%%%%%%%%%%%%%%%%%%%%%%%%%%%

%%%%%%%%%%%%%%%%%%%%%%%%%%%%%%%%%%%%%%%%%%%%%%%%%%%%%%%%%%%%%%%%%%%%%%%%%%%%%%%%%%%%%%%%%%%%%%%%%%%%%%%%%%%%%%%%%%%%%%%%%%%%%%%%%%%%%%%%%%%%%%%%%%%%%%%%%%%%%%%%%%%%%%%%%%%%%%%%%%%%%%%%%%%

Lastly, we are able to apply the analytic model developed here to both explain previous observations that pristine SWCNTs saturate at surprisingly low incident fluence, as well as exploit the nonlinear photoluminescence signal to extract the single SWCNT optical absorption. 
Even if the linear PL action cross-section ($\eta_{\rm{QE}}\times \sigma_{\rm{ab}}$) is well quantified, it is challenging to decouple optical absorption from quantum efficiency. 
This is a standard problem in nanostructure science.
With fluences sufficiently high to cause EEA, the optical absorption cross-section can be extracted since it is directly linked to injected exciton number through the fluence, $N=N_{\rm{c}}\times \sigma_{\rm{ab}} \times Fluence$, where $N_{\rm{c}}$ is the number of carbon atoms in the SWCNT.
Several experiments have observed that SWCNT PL saturates with increasing excitation fluence, but the saturation fluence between samples is orders of magnitude different \cite{Murakami_PRL09,Xiao_PRL10}.     
To verify diffusion as the main mechanism for exciton transport, this discrepancy must be reasonably accounted for given the variations between samples.
We simulate the PL saturation observed in PL studies using typical parameters for SWCNT samples for suspended and encapsulated bulk samples  (Figure \ref{fig:PPD} (c) where $PL=N_{0}\times\eta_{\rm{QE}}$).  
For encapsulated samples, a length of 610 nm, a diffusivity of 4 cm$^{2}/$s, and a relaxation rate of 50 ns$^{-1}$ (equivalent to a 90 nm diffusion length) are used. 
The corresponding parameters used for air-suspended samples are: 4.6 $\mu$m, 180 cm$^{2}/$s, and 2.9 ns$^{-1}$, which is are the parameters of one of the suspended SWCNTs studied.  
To quantify the saturation density for comparison purposes, we define the saturation point ($N_{\rm{sat}}$) as shown in (Figure \ref{fig:PPD} (c), with a simple model where $PL(N_{0})\propto 1-exp(-N_{0}/N_{\rm{sat}})$. 
Figure \ref{fig:PPD} (c) shows an order of magnitude difference in the exciton density at saturation.
An even larger ratio would be expected from the use of inhomogeneous excitation and randomly orientated SWCNTs as is often the case for ensemble studies of encapsulated SWCNTs\cite{Xiao_SPIE09}.

To extract the absorption cross-section, the diffusivity and decay rate are found for each SWCNT using FEC, and the length is found using PL spatial mapping.
Figure \ref{fig:PPD} (d) shows a fit of the experimental data for a single 4.6 $\mu$m (9,8) air-suspended SWCNT (from \citet{Xiao_PRL10}); the extracted absorption cross-section is $\sigma_{\rm{ab}}$ = 2.8 $\pm$ 0.1 $\times$10$^{-17}$ cm$^{2}$/atom, which is consistent with the highest reported value for SWCNTs\cite{Joh_NatNano11,Schoppler_JPhysChemC11}. 
The SWCNTs studied had quantum efficiencies from 5-11 $\%$ and absorption cross-sections from 2.1-3.7 $\times$ 10$^{-17}$ cm$^{2}$/atom. 
This is about five times higher than graphene ($\sim$ 0.6 $\times$ 10$^{-17}$ cm$^{2}$/atom \cite{Bonaccorso_NatPhot10}) consistent with an enhanced oscillator strength arising from the reduced dimensionality of the SWCNT.  
While the absorption cross-section is a standard parameter in literature, it may be more useful to state the frequency-integrated absorption cross-section\cite{Eberly_Lasers88}, $\sigma_{0}=\int_{0}^{\infty}\sigma(\omega)d\omega$.
It should be noted that the $E_{22}$ absorption linewidth for these samples is narrow (FWHM = 44 meV) compared to ensemble samples, and the excitation light is pulsed (FWHM = 18 meV) and polarized parallel to the SWCNT for maximum coupling. Under these conditions $\sigma_{0} = 0.0016-0.0028$ cm$^{2}/\rm{s}\cdot$atom.
%%%%%%%%%%%%%%%%%%%%%%%%%%%%%%%%%%%%%%%%%%%%%%%%%%%%%%%%%%%%%%%%%%%%%%%%%%%%%%%%%%%%%%%%%%%%%%%%%%%%%%%%%%%%%%%%%%%%%%%%%%%%%%%%%%%%%%%%%%%%%%%%%%%%%%%%%%%%%%%%%%%%%%%%%%%%%%%%%%%%%%%%%%%%%%%%%%%%%%%%%%%%%%%%%%%%%%%

By formulating exciton population dynamics, we are able to properly capture the length dependence of end quenching and EEA, as verified by Monte Carlo simulations.
From this formulation, we have a robust method of extracting the optical absorption cross-section, intrinsic lifetime, and diffusivity for SWCNTs using PL saturation and FEC experiments, even for a single SWCNT.
We observe that air-suspended SWCNTs are ``pristine'' with the highest optical absorption of any sample and extremely long intrinsic diffusion length, spanning several microns.
This gives credibility to EEA enabled single photon sources with selectable frequency in the telecom band.
Lastly, knowledge of a sample's intrinsic properties will allow for engineering of optoelectronic properties such as length or diffusivity to maximize the quantum efficiency (for optical emission applications) or decay rate (for applications requiring ultrafast relaxation such as saturable absorbers).

\begin{acknowledgments}
We acknowledge nanotube sample preparation by the group including P. Finnie and J. Lefebvre at the Security and Disruptive Technologies Portfolio, National Research Council Canada.
We thank M. Abdelqader, M. Dignam, S. Hughes, Y. Kato, and C. Van Vlack for helpful discussions.
This work is funded by the Natural Sciences and Engineering Research
Council of Canada, the Canadian Foundation for Innovation, and the
Ministry of Economic Development and Innovation (Ontario).
\end{acknowledgments}

%\bibliography{diffusionbib}

\begin{thebibliography}{40}%
\makeatletter
\providecommand \@ifxundefined [1]{%
 \@ifx{#1\undefined}
}%
\providecommand \@ifnum [1]{%
 \ifnum #1\expandafter \@firstoftwo
 \else \expandafter \@secondoftwo
 \fi
}%
\providecommand \@ifx [1]{%
 \ifx #1\expandafter \@firstoftwo
 \else \expandafter \@secondoftwo
 \fi
}%
\providecommand \natexlab [1]{#1}%
\providecommand \enquote  [1]{``#1''}%
\providecommand \bibnamefont  [1]{#1}%
\providecommand \bibfnamefont [1]{#1}%
\providecommand \citenamefont [1]{#1}%
\providecommand \href@noop [0]{\@secondoftwo}%
\providecommand \href [0]{\begingroup \@sanitize@url \@href}%
\providecommand \@href[1]{\@@startlink{#1}\@@href}%
\providecommand \@@href[1]{\endgroup#1\@@endlink}%
\providecommand \@sanitize@url [0]{\catcode `\\12\catcode `\$12\catcode
  `\&12\catcode `\#12\catcode `\^12\catcode `\_12\catcode `\%12\relax}%
\providecommand \@@startlink[1]{}%
\providecommand \@@endlink[0]{}%
\providecommand \url  [0]{\begingroup\@sanitize@url \@url }%
\providecommand \@url [1]{\endgroup\@href {#1}{\urlprefix }}%
\providecommand \urlprefix  [0]{URL }%
\providecommand \Eprint [0]{\href }%
\providecommand \doibase [0]{http://dx.doi.org/}%
\providecommand \selectlanguage [0]{\@gobble}%
\providecommand \bibinfo  [0]{\@secondoftwo}%
\providecommand \bibfield  [0]{\@secondoftwo}%
\providecommand \translation [1]{[#1]}%
\providecommand \BibitemOpen [0]{}%
\providecommand \bibitemStop [0]{}%
\providecommand \bibitemNoStop [0]{.\EOS\space}%
\providecommand \EOS [0]{\spacefactor3000\relax}%
\providecommand \BibitemShut  [1]{\csname bibitem#1\endcsname}%
\let\auto@bib@innerbib\@empty
%</preamble>
\bibitem [{\citenamefont {Perebeinos}\ \emph {et~al.}(2004)\citenamefont
  {Perebeinos}, \citenamefont {Tersoff},\ and\ \citenamefont
  {Avouris}}]{Perebeinos_PRL04}%
  \BibitemOpen
  \bibfield  {author} {\bibinfo {author} {\bibfnamefont {V.}~\bibnamefont
  {Perebeinos}}, \bibinfo {author} {\bibfnamefont {J.}~\bibnamefont {Tersoff}},
  \ and\ \bibinfo {author} {\bibfnamefont {P.}~\bibnamefont {Avouris}},\ }\href
  {\doibase 10.1103/PhysRevLett.92.257402} {\bibfield  {journal} {\bibinfo
  {journal} {Phys. Rev. Lett.}\ }\textbf {\bibinfo {volume} {92}},\ \bibinfo
  {pages} {257402} (\bibinfo {year} {2004})}\BibitemShut {NoStop}%
\bibitem [{\citenamefont {Wang}\ \emph {et~al.}(2005)\citenamefont {Wang},
  \citenamefont {Dukovic}, \citenamefont {Brus},\ and\ \citenamefont
  {Heinz}}]{Wang_Sci05}%
  \BibitemOpen
  \bibfield  {author} {\bibinfo {author} {\bibfnamefont {F.}~\bibnamefont
  {Wang}}, \bibinfo {author} {\bibfnamefont {G.}~\bibnamefont {Dukovic}},
  \bibinfo {author} {\bibfnamefont {L.~E.}\ \bibnamefont {Brus}}, \ and\
  \bibinfo {author} {\bibfnamefont {T.~F.}\ \bibnamefont {Heinz}},\ }\href@noop
  {} {\bibfield  {journal} {\bibinfo  {journal} {Science}\ }\textbf {\bibinfo
  {volume} {308}},\ \bibinfo {pages} {838} (\bibinfo {year}
  {2005})}\BibitemShut {NoStop}%
\bibitem [{\citenamefont {Cognet}\ \emph {et~al.}(2007)\citenamefont {Cognet},
  \citenamefont {Tsyboulski}, \citenamefont {Rocha}, \citenamefont {Doyle},
  \citenamefont {Tour},\ and\ \citenamefont {Weisman}}]{Cognet_Sci07}%
  \BibitemOpen
  \bibfield  {author} {\bibinfo {author} {\bibfnamefont {L.}~\bibnamefont
  {Cognet}}, \bibinfo {author} {\bibfnamefont {D.}~\bibnamefont {Tsyboulski}},
  \bibinfo {author} {\bibfnamefont {J.}~\bibnamefont {Rocha}}, \bibinfo
  {author} {\bibfnamefont {C.}~\bibnamefont {Doyle}}, \bibinfo {author}
  {\bibfnamefont {J.}~\bibnamefont {Tour}}, \ and\ \bibinfo {author}
  {\bibfnamefont {R.}~\bibnamefont {Weisman}},\ }\href@noop {} {\bibfield
  {journal} {\bibinfo  {journal} {Science}\ }\textbf {\bibinfo {volume}
  {316}},\ \bibinfo {pages} {1465} (\bibinfo {year} {2007})}\BibitemShut
  {NoStop}%
\bibitem [{\citenamefont {Liu}\ and\ \citenamefont
  {Xiao}(2011)}]{Liu_JPhysChemC11}%
  \BibitemOpen
  \bibfield  {author} {\bibinfo {author} {\bibfnamefont {T.}~\bibnamefont
  {Liu}}\ and\ \bibinfo {author} {\bibfnamefont {Z.}~\bibnamefont {Xiao}},\
  }\href {\doibase 10.1021/jp205458t} {\bibfield  {journal} {\bibinfo
  {journal} {J. Phys. Chem. C}\ }\textbf {\bibinfo {volume} {115}},\ \bibinfo
  {pages} {16920} (\bibinfo {year} {2011})}\BibitemShut {NoStop}%
\bibitem [{\citenamefont {Ma}\ \emph {et~al.}(2005)\citenamefont {Ma},
  \citenamefont {Valkunas}, \citenamefont {Dexheimer}, \citenamefont
  {Bachilo},\ and\ \citenamefont {Fleming}}]{Ma_PRL05}%
  \BibitemOpen
  \bibfield  {author} {\bibinfo {author} {\bibfnamefont {Y.-Z.}\ \bibnamefont
  {Ma}}, \bibinfo {author} {\bibfnamefont {L.}~\bibnamefont {Valkunas}},
  \bibinfo {author} {\bibfnamefont {S.~L.}\ \bibnamefont {Dexheimer}}, \bibinfo
  {author} {\bibfnamefont {S.~M.}\ \bibnamefont {Bachilo}}, \ and\ \bibinfo
  {author} {\bibfnamefont {G.~R.}\ \bibnamefont {Fleming}},\ }\href {\doibase
  10.1103/PhysRevLett.94.157402} {\bibfield  {journal} {\bibinfo  {journal}
  {Phys. Rev. Lett.}\ }\textbf {\bibinfo {volume} {94}},\ \bibinfo {pages}
  {157402} (\bibinfo {year} {2005})}\BibitemShut {NoStop}%
\bibitem [{\citenamefont {Huang}\ and\ \citenamefont
  {Krauss}(2006)}]{Huang_PRL06}%
  \BibitemOpen
  \bibfield  {author} {\bibinfo {author} {\bibfnamefont {L.}~\bibnamefont
  {Huang}}\ and\ \bibinfo {author} {\bibfnamefont {T.D.}~\bibnamefont {Krauss}},\
  }\href@noop {} {\bibfield  {journal} {\bibinfo  {journal} {Phys. Rev. Lett.}\
  }\textbf {\bibinfo {volume} {96}},\ \bibinfo {pages} {057407} (\bibinfo {year}
  {2006})}\BibitemShut {NoStop}%
\bibitem [{\citenamefont {Barone}\ \emph {et~al.}(2004)\citenamefont {Barone},
  \citenamefont {Baik}, \citenamefont {Heller},\ and\ \citenamefont
  {Strano}}]{Barone_NatMat04}%
  \BibitemOpen
  \bibfield  {author} {\bibinfo {author} {\bibfnamefont {P.}~\bibnamefont
  {Barone}}, \bibinfo {author} {\bibfnamefont {S.}~\bibnamefont {Baik}},
  \bibinfo {author} {\bibfnamefont {D.}~\bibnamefont {Heller}}, \ and\ \bibinfo
  {author} {\bibfnamefont {M.}~\bibnamefont {Strano}},\ }\href@noop {}
  {\bibfield  {journal} {\bibinfo  {journal} {Nat. Mater.}\ }\textbf {\bibinfo
  {volume} {4}},\ \bibinfo {pages} {86} (\bibinfo {year} {2004})}\BibitemShut
  {NoStop}%
\bibitem [{\citenamefont {Kim}\ \emph {et~al.}(2009)\citenamefont {Kim},
  \citenamefont {Heller}, \citenamefont {Jin}, \citenamefont {Barone},
  \citenamefont {Song}, \citenamefont {Zhang}, \citenamefont {Trudel},
  \citenamefont {Wogan}, \citenamefont {Tannenbaum},\ and\ \citenamefont
  {Strano}}]{Kim_NatureChem09}%
  \BibitemOpen
  \bibfield  {author} {\bibinfo {author} {\bibfnamefont {J.}~\bibnamefont
  {Kim}}, \bibinfo {author} {\bibfnamefont {D.}~\bibnamefont {Heller}},
  \bibinfo {author} {\bibfnamefont {H.}~\bibnamefont {Jin}}, \bibinfo {author}
  {\bibfnamefont {P.}~\bibnamefont {Barone}}, \bibinfo {author} {\bibfnamefont
  {C.}~\bibnamefont {Song}}, \bibinfo {author} {\bibfnamefont {J.}~\bibnamefont
  {Zhang}}, \bibinfo {author} {\bibfnamefont {L.}~\bibnamefont {Trudel}},
  \bibinfo {author} {\bibfnamefont {G.}~\bibnamefont {Wogan}}, \bibinfo
  {author} {\bibfnamefont {S.}~\bibnamefont {Tannenbaum}}, \ and\ \bibinfo
  {author} {\bibfnamefont {M.}~\bibnamefont {Strano}},\ }\href@noop {}
  {\bibfield  {journal} {\bibinfo  {journal} {Nat. Chem.}\ }\textbf {\bibinfo
  {volume} {1}},\ \bibinfo {pages} {473} (\bibinfo {year} {2009})}\BibitemShut
  {NoStop}%
\bibitem [{\citenamefont {Ulissi}\ \emph {et~al.}(2011)\citenamefont {Ulissi},
  \citenamefont {Zhang}, \citenamefont {Boghossian}, \citenamefont {Reuel},
  \citenamefont {Shimizu}, \citenamefont {Braatz},\ and\ \citenamefont
  {Strano}}]{Ulissi_JPhysChemLett11}%
  \BibitemOpen
  \bibfield  {author} {\bibinfo {author} {\bibfnamefont {Z.~W.}\ \bibnamefont
  {Ulissi}}, \bibinfo {author} {\bibfnamefont {J.}~\bibnamefont {Zhang}},
  \bibinfo {author} {\bibfnamefont {A.~A.}\ \bibnamefont {Boghossian}},
  \bibinfo {author} {\bibfnamefont {N.~F.}\ \bibnamefont {Reuel}}, \bibinfo
  {author} {\bibfnamefont {S.~F.~E.}\ \bibnamefont {Shimizu}}, \bibinfo
  {author} {\bibfnamefont {R.~D.}\ \bibnamefont {Braatz}}, \ and\ \bibinfo
  {author} {\bibfnamefont {M.~S.}\ \bibnamefont {Strano}},\ }\href {\doibase
  10.1021/jz200572b} {\bibfield  {journal} {\bibinfo  {journal} {J. Phys. Chem.
  Lett.}\ }\textbf {\bibinfo {volume} {2}},\ \bibinfo {pages} {1690} (\bibinfo
  {year} {2011})}\BibitemShut {NoStop}%
\bibitem [{\citenamefont {H\"ogele}\ \emph {et~al.}(2008)\citenamefont
  {H\"ogele}, \citenamefont {Galland}, \citenamefont {Winger},\ and\
  \citenamefont {Imamo\ifmmode~\breve{g}\else \u{g}\fi{}lu}}]{Hogele_PRL08}%
  \BibitemOpen
  \bibfield  {author} {\bibinfo {author} {\bibfnamefont {A.}~\bibnamefont
  {H\"ogele}}, \bibinfo {author} {\bibfnamefont {C.}~\bibnamefont {Galland}},
  \bibinfo {author} {\bibfnamefont {M.}~\bibnamefont {Winger}}, \ and\ \bibinfo
  {author} {\bibfnamefont {A.}~\bibnamefont {Imamo\ifmmode~\breve{g}\else
  \u{g}\fi{}lu}},\ }\href {\doibase 10.1103/PhysRevLett.100.217401} {\bibfield
  {journal} {\bibinfo  {journal} {Phys. Rev. Lett.}\ }\textbf {\bibinfo
  {volume} {100}},\ \bibinfo {pages} {217401} (\bibinfo {year}
  {2008})}\BibitemShut {NoStop}%
\bibitem [{\citenamefont {Harrah}\ \emph {et~al.}(2011)\citenamefont {Harrah},
  \citenamefont {Schneck}, \citenamefont {Green}, \citenamefont {Hersam},
  \citenamefont {Ziegler},\ and\ \citenamefont {Swan}}]{Harrah_Nano11}%
  \BibitemOpen
  \bibfield  {author} {\bibinfo {author} {\bibfnamefont {D.~M.}\ \bibnamefont
  {Harrah}}, \bibinfo {author} {\bibfnamefont {J.~R.}\ \bibnamefont {Schneck}},
  \bibinfo {author} {\bibfnamefont {A.~A.}\ \bibnamefont {Green}}, \bibinfo
  {author} {\bibfnamefont {M.~C.}\ \bibnamefont {Hersam}}, \bibinfo {author}
  {\bibfnamefont {L.~D.}\ \bibnamefont {Ziegler}}, \ and\ \bibinfo {author}
  {\bibfnamefont {A.~K.}\ \bibnamefont {Swan}},\ }\href {\doibase
  10.1021/nn203604v} {\bibfield  {journal} {\bibinfo  {journal} {ACS Nano}\
  }\textbf {\bibinfo {volume} {5}},\ \bibinfo {pages} {9898} (\bibinfo {year}
  {2011})}\BibitemShut {NoStop}%
\bibitem [{\citenamefont {Joh}\ \emph {et~al.}(2011)\citenamefont {Joh},
  \citenamefont {Kinder}, \citenamefont {Herman}, \citenamefont {Ju},
  \citenamefont {Segal}, \citenamefont {Chan},\ and\ \citenamefont
  {Park}}]{Joh_NatNano11}%
  \BibitemOpen
  \bibfield  {author} {\bibinfo {author} {\bibfnamefont {D.~Y.}\ \bibnamefont
  {Joh}}, \bibinfo {author} {\bibfnamefont {J.}~\bibnamefont {Kinder}},
  \bibinfo {author} {\bibfnamefont {L.~H.}\ \bibnamefont {Herman}}, \bibinfo
  {author} {\bibfnamefont {S.-Y.}\ \bibnamefont {Ju}}, \bibinfo {author}
  {\bibfnamefont {J.~N.}\ \bibnamefont {Segal}, \bibfnamefont {Michael
  A.~Johnson}}, \bibinfo {author} {\bibfnamefont {G.~K.-L.}\ \bibnamefont
  {Chan}}, \ and\ \bibinfo {author} {\bibfnamefont {J.}~\bibnamefont {Park}},\
  }\href {\doibase 10.1038/nnano.2010.248} {\bibfield  {journal} {\bibinfo
  {journal} {Nat. Nanotechnol.}\ }\textbf {\bibinfo {volume} {6}},\ \bibinfo
  {pages} {51} (\bibinfo {year} {2011})}\BibitemShut {NoStop}%
\bibitem [{\citenamefont {SchoÌˆppler}\ \emph {et~al.}(2011)\citenamefont
  {SchoÌˆppler}, \citenamefont {Mann}, \citenamefont {Hain}, \citenamefont
  {Neubauer}, \citenamefont {Privitera}, \citenamefont {Bonaccorso},
  \citenamefont {Chu}, \citenamefont {Ferrari},\ and\ \citenamefont
  {Hertel}}]{Schoppler_JPhysChemC11}%
  \BibitemOpen
  \bibfield  {author} {\bibinfo {author} {\bibfnamefont {F.}~\bibnamefont
  {SchoÌˆppler}}, \bibinfo {author} {\bibfnamefont {C.}~\bibnamefont {Mann}},
  \bibinfo {author} {\bibfnamefont {T.~C.}\ \bibnamefont {Hain}}, \bibinfo
  {author} {\bibfnamefont {F.~M.}\ \bibnamefont {Neubauer}}, \bibinfo {author}
  {\bibfnamefont {G.}~\bibnamefont {Privitera}}, \bibinfo {author}
  {\bibfnamefont {F.}~\bibnamefont {Bonaccorso}}, \bibinfo {author}
  {\bibfnamefont {D.}~\bibnamefont {Chu}}, \bibinfo {author} {\bibfnamefont
  {A.~C.}\ \bibnamefont {Ferrari}}, \ and\ \bibinfo {author} {\bibfnamefont
  {T.}~\bibnamefont {Hertel}},\ }\href {\doibase 10.1021/jp205289h} {\bibfield
  {journal} {\bibinfo  {journal} {J. Phys. Chem. C}\ }\textbf {\bibinfo
  {volume} {115}},\ \bibinfo {pages} {14682} (\bibinfo {year}
  {2011})}\BibitemShut {NoStop}%
\bibitem [{\citenamefont {Berciaud}\ \emph {et~al.}(2008)\citenamefont
  {Berciaud}, \citenamefont {Cognet},\ and\ \citenamefont
  {Lounis}}]{Berciaud_PRL08}%
  \BibitemOpen
  \bibfield  {author} {\bibinfo {author} {\bibfnamefont {S.}~\bibnamefont
  {Berciaud}}, \bibinfo {author} {\bibfnamefont {L.}~\bibnamefont {Cognet}}, \
  and\ \bibinfo {author} {\bibfnamefont {B.}~\bibnamefont {Lounis}},\
  }\href@noop {} {\bibfield  {journal} {\bibinfo  {journal} {Phys. Rev. Lett.}\
  }\textbf {\bibinfo {volume} {101}},\ \bibinfo {pages} {077402} (\bibinfo
  {year} {2008})}\BibitemShut {NoStop}%
\bibitem [{\citenamefont {Ruzicka}\ \emph {et~al.}(2012)\citenamefont
  {Ruzicka}, \citenamefont {Wang}, \citenamefont {Lohrman}, \citenamefont
  {Ren},\ and\ \citenamefont {Zhao}}]{Ruzicka_PRB12}%
  \BibitemOpen
  \bibfield  {author} {\bibinfo {author} {\bibfnamefont {B.~A.}\ \bibnamefont
  {Ruzicka}}, \bibinfo {author} {\bibfnamefont {R.}~\bibnamefont {Wang}},
  \bibinfo {author} {\bibfnamefont {J.}~\bibnamefont {Lohrman}}, \bibinfo
  {author} {\bibfnamefont {S.}~\bibnamefont {Ren}}, \ and\ \bibinfo {author}
  {\bibfnamefont {H.}~\bibnamefont {Zhao}},\ }\href {\doibase
  10.1103/PhysRevB.86.205417} {\bibfield  {journal} {\bibinfo  {journal} {Phys.
  Rev. B}\ }\textbf {\bibinfo {volume} {86}},\ \bibinfo {pages} {205417}
  (\bibinfo {year} {2012})}\BibitemShut {NoStop}%
\bibitem [{\citenamefont {Harrah}\ and\ \citenamefont
  {Swan}(2011)}]{Harrah_Nano10}%
  \BibitemOpen
  \bibfield  {author} {\bibinfo {author} {\bibfnamefont {D.~M.}\ \bibnamefont
  {Harrah}}\ and\ \bibinfo {author} {\bibfnamefont {A.~K.}\ \bibnamefont
  {Swan}},\ }\href {\doibase 10.1021/nn1031214} {\bibfield  {journal} {\bibinfo
   {journal} {ACS Nano}\ }\textbf {\bibinfo {volume} {5}},\ \bibinfo {pages}
  {647} (\bibinfo {year} {2011})}\BibitemShut {NoStop}%
\bibitem [{\citenamefont {Murakami}\ and\ \citenamefont
  {Kono}(2009)}]{Murakami_PRL09}%
  \BibitemOpen
  \bibfield  {author} {\bibinfo {author} {\bibfnamefont {Y.}~\bibnamefont
  {Murakami}}\ and\ \bibinfo {author} {\bibfnamefont {J.}~\bibnamefont
  {Kono}},\ }\href {\doibase 10.1103/PhysRevLett.102.037401} {\bibfield
  {journal} {\bibinfo  {journal} {Phys. Rev. Lett.}\ }\textbf {\bibinfo
  {volume} {102}},\ \bibinfo {pages} {037401} (\bibinfo {year}
  {2009})}\BibitemShut {NoStop}%
\bibitem [{\citenamefont {Xiao}\ \emph {et~al.}(2010)\citenamefont {Xiao},
  \citenamefont {Nhan}, \citenamefont {Wilson},\ and\ \citenamefont
  {Fraser}}]{Xiao_PRL10}%
  \BibitemOpen
  \bibfield  {author} {\bibinfo {author} {\bibfnamefont {Y.-F.}\ \bibnamefont
  {Xiao}}, \bibinfo {author} {\bibfnamefont {T.~Q.}\ \bibnamefont {Nhan}},
  \bibinfo {author} {\bibfnamefont {M.~W.~B.}\ \bibnamefont {Wilson}}, \ and\
  \bibinfo {author} {\bibfnamefont {J.~M.}\ \bibnamefont {Fraser}},\ }\href
  {\doibase 10.1103/PhysRevLett.104.017401} {\bibfield  {journal} {\bibinfo
  {journal} {Phys. Rev. Lett.}\ }\textbf {\bibinfo {volume} {104}},\ \bibinfo
  {pages} {017401} (\bibinfo {year} {2010})}\BibitemShut {NoStop}%
\bibitem [{\citenamefont {Moritsubo}\ \emph {et~al.}(2010)\citenamefont
  {Moritsubo}, \citenamefont {Murai}, \citenamefont {Shimada}, \citenamefont
  {Murakami}, \citenamefont {Chiashi}, \citenamefont {Maruyama},\ and\
  \citenamefont {Kato}}]{Moritsubo_PRL10}%
  \BibitemOpen
  \bibfield  {author} {\bibinfo {author} {\bibfnamefont {S.}~\bibnamefont
  {Moritsubo}}, \bibinfo {author} {\bibfnamefont {T.}~\bibnamefont {Murai}},
  \bibinfo {author} {\bibfnamefont {T.}~\bibnamefont {Shimada}}, \bibinfo
  {author} {\bibfnamefont {Y.}~\bibnamefont {Murakami}}, \bibinfo {author}
  {\bibfnamefont {S.}~\bibnamefont {Chiashi}}, \bibinfo {author} {\bibfnamefont
  {S.}~\bibnamefont {Maruyama}}, \ and\ \bibinfo {author} {\bibfnamefont
  {Y.~K.}\ \bibnamefont {Kato}},\ }\href {\doibase
  10.1103/PhysRevLett.104.247402} {\bibfield  {journal} {\bibinfo  {journal}
  {Phys. Rev. Lett.}\ }\textbf {\bibinfo {volume} {104}},\ \bibinfo {pages}
  {247402} (\bibinfo {year} {2010})}\BibitemShut {NoStop}%
\bibitem [{\citenamefont {Xie}\ \emph {et~al.}(2012)\citenamefont {Xie},
  \citenamefont {Inaba}, \citenamefont {Sugiyama},\ and\ \citenamefont
  {Homma}}]{Xie_PRB12}%
  \BibitemOpen
  \bibfield  {author} {\bibinfo {author} {\bibfnamefont {J.}~\bibnamefont
  {Xie}}, \bibinfo {author} {\bibfnamefont {T.}~\bibnamefont {Inaba}}, \bibinfo
  {author} {\bibfnamefont {R.}~\bibnamefont {Sugiyama}}, \ and\ \bibinfo
  {author} {\bibfnamefont {Y.}~\bibnamefont {Homma}},\ }\href {\doibase
  10.1103/PhysRevB.85.085434} {\bibfield  {journal} {\bibinfo  {journal} {Phys.
  Rev. B}\ }\textbf {\bibinfo {volume} {85}},\ \bibinfo {pages} {085434}
  (\bibinfo {year} {2012})}\BibitemShut {NoStop}%
\bibitem [{\citenamefont {Rajan}\ \emph {et~al.}(2008)\citenamefont {Rajan},
  \citenamefont {Strano}, \citenamefont {Heller}, \citenamefont {Hertel},\ and\
  \citenamefont {Schulten}}]{Rajan_PhysChemB08}%
  \BibitemOpen
  \bibfield  {author} {\bibinfo {author} {\bibfnamefont {A.}~\bibnamefont
  {Rajan}}, \bibinfo {author} {\bibfnamefont {M.~S.}\ \bibnamefont {Strano}},
  \bibinfo {author} {\bibfnamefont {D.~A.}\ \bibnamefont {Heller}}, \bibinfo
  {author} {\bibfnamefont {T.}~\bibnamefont {Hertel}}, \ and\ \bibinfo {author}
  {\bibfnamefont {K.}~\bibnamefont {Schulten}},\ }\href {\doibase
  10.1021/jp077144l} {\bibfield  {journal} {\bibinfo  {journal} {J. Phys. Chem.
  B}\ }\textbf {\bibinfo {volume} {112}},\ \bibinfo {pages} {6211} (\bibinfo
  {year} {2008})},\ \bibinfo {note} {pMID: 18327930}\BibitemShut {NoStop}%
\bibitem [{\citenamefont {Hertel}\ \emph {et~al.}(2010)\citenamefont {Hertel},
  \citenamefont {Himmelein}, \citenamefont {Ackermann}, \citenamefont {Stich},\
  and\ \citenamefont {Crochet}}]{Hertel_Nano10}%
  \BibitemOpen
  \bibfield  {author} {\bibinfo {author} {\bibfnamefont {T.}~\bibnamefont
  {Hertel}}, \bibinfo {author} {\bibfnamefont {S.}~\bibnamefont {Himmelein}},
  \bibinfo {author} {\bibfnamefont {T.}~\bibnamefont {Ackermann}}, \bibinfo
  {author} {\bibfnamefont {D.}~\bibnamefont {Stich}}, \ and\ \bibinfo {author}
  {\bibfnamefont {J.}~\bibnamefont {Crochet}},\ }\href {\doibase
  10.1021/nn101612b} {\bibfield  {journal} {\bibinfo  {journal} {ACS Nano}\
  }\textbf {\bibinfo {volume} {4}},\ \bibinfo {pages} {7161} (\bibinfo {year}
  {2010})}\BibitemShut {NoStop}%
\bibitem [{\citenamefont {Siitonen}\ \emph {et~al.}(2010)\citenamefont
  {Siitonen}, \citenamefont {Tsyboulski}, \citenamefont {Bachilo},\ and\
  \citenamefont {Weisman}}]{Siitonen_NL10}%
  \BibitemOpen
  \bibfield  {author} {\bibinfo {author} {\bibfnamefont {A.~J.}\ \bibnamefont
  {Siitonen}}, \bibinfo {author} {\bibfnamefont {D.~A.}\ \bibnamefont
  {Tsyboulski}}, \bibinfo {author} {\bibfnamefont {S.~M.}\ \bibnamefont
  {Bachilo}}, \ and\ \bibinfo {author} {\bibfnamefont {R.~B.}\ \bibnamefont
  {Weisman}},\ }\href {\doibase 10.1021/nl9039845} {\bibfield  {journal}
  {\bibinfo  {journal} {Nano Lett.}\ }\textbf {\bibinfo {volume} {10}},\
  \bibinfo {pages} {1595} (\bibinfo {year} {2010})}\BibitemShut {NoStop}%
\bibitem [{\citenamefont {Siitonen}\ \emph {et~al.}(2012)\citenamefont
  {Siitonen}, \citenamefont {Bachilo}, \citenamefont {Tsyboulski},\ and\
  \citenamefont {Weisman}}]{Siitonen_NL12}%
  \BibitemOpen
  \bibfield  {author} {\bibinfo {author} {\bibfnamefont {A.~J.}\ \bibnamefont
  {Siitonen}}, \bibinfo {author} {\bibfnamefont {S.~M.}\ \bibnamefont
  {Bachilo}}, \bibinfo {author} {\bibfnamefont {D.~A.}\ \bibnamefont
  {Tsyboulski}}, \ and\ \bibinfo {author} {\bibfnamefont {R.~B.}\ \bibnamefont
  {Weisman}},\ }\href {\doibase 10.1021/nl2028238} {\bibfield  {journal}
  {\bibinfo  {journal} {Nano Lett.}\ }\textbf {\bibinfo {volume} {12}},\
  \bibinfo {pages} {33} (\bibinfo {year} {2012})}\BibitemShut {NoStop}%
\bibitem [{\citenamefont {Naqvi}\ \emph {et~al.}(1982)\citenamefont {Naqvi},
  \citenamefont {Mork},\ and\ \citenamefont {Waldenstr\o{}m}}]{Naqvi_PRL82}%
  \BibitemOpen
  \bibfield  {author} {\bibinfo {author} {\bibfnamefont {K.~R.}\ \bibnamefont
  {Naqvi}}, \bibinfo {author} {\bibfnamefont {K.~J.}\ \bibnamefont {Mork}}, \
  and\ \bibinfo {author} {\bibfnamefont {S.}~\bibnamefont {Waldenstr\o{}m}},\
  }\href {\doibase 10.1103/PhysRevLett.49.304} {\bibfield  {journal} {\bibinfo
  {journal} {Phys. Rev. Lett.}\ }\textbf {\bibinfo {volume} {49}},\ \bibinfo
  {pages} {304} (\bibinfo {year} {1982})}\BibitemShut {NoStop}%
\bibitem [{\citenamefont {Harris}(1980{\natexlab{a}})}]{Harris_JPhysA80}%
  \BibitemOpen
  \bibfield  {author} {\bibinfo {author} {\bibfnamefont {S.}~\bibnamefont
  {Harris}},\ }\href@noop {} {\bibfield  {journal} {\bibinfo  {journal} {J.
  Phys. A-Math Gen}\ }\textbf {\bibinfo {volume} {13}},\ \bibinfo {pages}
  {2149} (\bibinfo {year} {1980}{\natexlab{a}})}\BibitemShut {NoStop}%
\bibitem [{\citenamefont {Harris}(1980{\natexlab{b}})}]{Harris_JCHEMPHYS80}%
  \BibitemOpen
  \bibfield  {author} {\bibinfo {author} {\bibfnamefont {S.}~\bibnamefont
  {Harris}},\ }\href {\doibase 10.1063/1.439411} {\bibfield  {journal}
  {\bibinfo  {journal} {J. Chem. Phys.}\ }\textbf {\bibinfo {volume} {72}},\
  \bibinfo {pages} {2659} (\bibinfo {year} {1980}{\natexlab{b}})}\BibitemShut
  {NoStop}%
\bibitem [{\citenamefont {Harris}(1987)}]{Harris_PRA87}%
  \BibitemOpen
  \bibfield  {author} {\bibinfo {author} {\bibfnamefont {S.}~\bibnamefont
  {Harris}},\ }\href {\doibase 10.1103/PhysRevA.36.3392} {\bibfield  {journal}
  {\bibinfo  {journal} {Phys. Rev. A}\ }\textbf {\bibinfo {volume} {36}},\
  \bibinfo {pages} {3392} (\bibinfo {year} {1987})}\BibitemShut {NoStop}%
\bibitem [{\citenamefont {Bray}\ and\ \citenamefont
  {Blythe}(2002)}]{Bray_PRL02}%
  \BibitemOpen
  \bibfield  {author} {\bibinfo {author} {\bibfnamefont {A.~J.}\ \bibnamefont
  {Bray}}\ and\ \bibinfo {author} {\bibfnamefont {R.~A.}\ \bibnamefont
  {Blythe}},\ }\href {\doibase 10.1103/PhysRevLett.89.150601} {\bibfield
  {journal} {\bibinfo  {journal} {Phys. Rev. Lett.}\ }\textbf {\bibinfo
  {volume} {89}},\ \bibinfo {pages} {150601} (\bibinfo {year}
  {2002})}\BibitemShut {NoStop}%
\bibitem [{\citenamefont {Miyauchi}\ \emph {et~al.}(2009)\citenamefont
  {Miyauchi}, \citenamefont {Matsuda},\ and\ \citenamefont
  {Kanemitsu}}]{MiyauchiFEC_PRB09}%
  \BibitemOpen
  \bibfield  {author} {\bibinfo {author} {\bibfnamefont {Y.}~\bibnamefont
  {Miyauchi}}, \bibinfo {author} {\bibfnamefont {K.}~\bibnamefont {Matsuda}}, \
  and\ \bibinfo {author} {\bibfnamefont {Y.}~\bibnamefont {Kanemitsu}},\ }\href
  {\doibase 10.1103/PhysRevB.80.235433} {\bibfield  {journal} {\bibinfo
  {journal} {Phys. Rev. B}\ }\textbf {\bibinfo {volume} {80}},\ \bibinfo
  {pages} {235433} (\bibinfo {year} {2009})}\BibitemShut {NoStop}%
\bibitem [{\citenamefont {Perebeinos}\ and\ \citenamefont
  {Avouris}(2008)}]{Perebeinos_PRL08}%
  \BibitemOpen
  \bibfield  {author} {\bibinfo {author} {\bibfnamefont {V.}~\bibnamefont
  {Perebeinos}}\ and\ \bibinfo {author} {\bibfnamefont {P.}~\bibnamefont
  {Avouris}},\ }\href {\doibase 10.1103/PhysRevLett.101.057401} {\bibfield
  {journal} {\bibinfo  {journal} {Phys. Rev. Lett.}\ }\textbf {\bibinfo
  {volume} {101}},\ \bibinfo {pages} {057401} (\bibinfo {year}
  {2008})}\BibitemShut {NoStop}%
\bibitem [{\citenamefont {Hagen}\ \emph {et~al.}(2005)\citenamefont {Hagen},
  \citenamefont {Steiner}, \citenamefont {Raschke}, \citenamefont {Lienau},
  \citenamefont {Hertel}, \citenamefont {Qian}, \citenamefont {Meixner},\ and\
  \citenamefont {Hartschuh}}]{Hagen_PRL05}%
  \BibitemOpen
  \bibfield  {author} {\bibinfo {author} {\bibfnamefont {A.}~\bibnamefont
  {Hagen}}, \bibinfo {author} {\bibfnamefont {M.}~\bibnamefont {Steiner}},
  \bibinfo {author} {\bibfnamefont {M.B.}~\bibnamefont {Raschke}}, \bibinfo
  {author} {\bibfnamefont {C.}~\bibnamefont {Lienau}}, \bibinfo {author}
  {\bibfnamefont {T.}~\bibnamefont {Hertel}}, \bibinfo {author} {\bibfnamefont
  {H.}~\bibnamefont {Qian}}, \bibinfo {author} {\bibfnamefont {A.J.}~\bibnamefont
  {Meixner}}, \ and\ \bibinfo {author} {\bibfnamefont {A.}~\bibnamefont
  {Hartschuh}},\ }\href@noop {} {\bibfield  {journal} {\bibinfo  {journal}
  {Phys. Rev. Lett.}\ }\textbf {\bibinfo {volume} {95}},\ \bibinfo {pages}
  {197401} (\bibinfo {year} {2005})}\BibitemShut {NoStop}%
\bibitem [{\citenamefont {Karatzas}\ and\ \citenamefont
  {Shreve}(1991)}]{Karatzas_BMSS91}%
  \BibitemOpen
  \bibfield  {author} {\bibinfo {author} {\bibfnamefont {I.}~\bibnamefont
  {Karatzas}}\ and\ \bibinfo {author} {\bibfnamefont {S.~E.}\ \bibnamefont
  {Shreve}},\ }\href@noop {} {\emph {\bibinfo {title} {Brownian Motion and
  Stochastic Calculus}}},\ \bibinfo {edition} {2nd}\ ed.\ (\bibinfo
  {publisher} {Springer},\ \bibinfo {address} {New York, New York},\ \bibinfo
  {year} {1991})\BibitemShut {NoStop}%
\bibitem [{\citenamefont {Miyauchi}\ \emph {et~al.}(2010)\citenamefont
  {Miyauchi}, \citenamefont {Matsuda}, \citenamefont {Yamamoto}, \citenamefont
  {Nakashima},\ and\ \citenamefont {Kanemitsu}}]{Miyauchi_JPhysChemC10}%
  \BibitemOpen
  \bibfield  {author} {\bibinfo {author} {\bibfnamefont {Y.}~\bibnamefont
  {Miyauchi}}, \bibinfo {author} {\bibfnamefont {K.}~\bibnamefont {Matsuda}},
  \bibinfo {author} {\bibfnamefont {Y.}~\bibnamefont {Yamamoto}}, \bibinfo
  {author} {\bibfnamefont {N.}~\bibnamefont {Nakashima}}, \ and\ \bibinfo
  {author} {\bibfnamefont {Y.}~\bibnamefont {Kanemitsu}},\ }\href {\doibase
  10.1021/jp1027492} {\bibfield  {journal} {\bibinfo  {journal} {J. Phys. Chem.
  C}\ }\textbf {\bibinfo {volume} {114}},\ \bibinfo {pages} {12905} (\bibinfo
  {year} {2010})}\BibitemShut {NoStop}%
\bibitem [{\citenamefont {Crochet}\ \emph {et~al.}(2012)\citenamefont
  {Crochet}, \citenamefont {Duque}, \citenamefont {Werner}, \citenamefont
  {Lounis}, \citenamefont {Cognet},\ and\ \citenamefont
  {Doorn}}]{Crochet_NL12}%
  \BibitemOpen
  \bibfield  {author} {\bibinfo {author} {\bibfnamefont {J.~J.}\ \bibnamefont
  {Crochet}}, \bibinfo {author} {\bibfnamefont {J.~G.}\ \bibnamefont {Duque}},
  \bibinfo {author} {\bibfnamefont {J.~H.}\ \bibnamefont {Werner}}, \bibinfo
  {author} {\bibfnamefont {B.}~\bibnamefont {Lounis}}, \bibinfo {author}
  {\bibfnamefont {L.}~\bibnamefont {Cognet}}, \ and\ \bibinfo {author}
  {\bibfnamefont {S.~K.}\ \bibnamefont {Doorn}},\ }\href {\doibase
  10.1021/nl301739d} {\bibfield  {journal} {\bibinfo  {journal} {Nano Lett.}\
  }\textbf {\bibinfo {volume} {12}},\ \bibinfo {pages} {5091} (\bibinfo {year}
  {2012})}\BibitemShut {NoStop}%
\bibitem [{\citenamefont {Tonks}(1936)}]{Tonks_PR36}%
  \BibitemOpen
  \bibfield  {author} {\bibinfo {author} {\bibfnamefont {L.}~\bibnamefont
  {Tonks}},\ }\href {\doibase 10.1103/PhysRev.50.955} {\bibfield  {journal}
  {\bibinfo  {journal} {Phys. Rev.}\ }\textbf {\bibinfo {volume} {50}},\
  \bibinfo {pages} {955} (\bibinfo {year} {1936})}\BibitemShut {NoStop}%
\bibitem [{\citenamefont {Srivastava}\ and\ \citenamefont
  {Kono}(2009)}]{Srivastava_PRB09}%
  \BibitemOpen
  \bibfield  {author} {\bibinfo {author} {\bibfnamefont {A.}~\bibnamefont
  {Srivastava}}\ and\ \bibinfo {author} {\bibfnamefont {J.}~\bibnamefont
  {Kono}},\ }\href {\doibase 10.1103/PhysRevB.79.205407} {\bibfield  {journal}
  {\bibinfo  {journal} {Phys. Rev. B}\ }\textbf {\bibinfo {volume} {79}},\
  \bibinfo {pages} {205407} (\bibinfo {year} {2009})}\BibitemShut {NoStop}%
\bibitem [{\citenamefont {Xiao}\ \emph {et~al.}(2009)\citenamefont {Xiao},
  \citenamefont {Nhan}, \citenamefont {Wilson},\ and\ \citenamefont
  {Fraser}}]{Xiao_SPIE09}%
  \BibitemOpen
  \bibfield  {author} {\bibinfo {author} {\bibfnamefont {Y.-F.}\ \bibnamefont
  {Xiao}}, \bibinfo {author} {\bibfnamefont {T.~Q.}\ \bibnamefont {Nhan}},
  \bibinfo {author} {\bibfnamefont {M.~W.~B.}\ \bibnamefont {Wilson}}, \ and\
  \bibinfo {author} {\bibfnamefont {J.~M.}\ \bibnamefont {Fraser}},\ }\href
  {\doibase 10.1117/12.809467} {\bibfield  {journal} {\bibinfo  {journal}
  {SPIE}\ ,\ \bibinfo {pages} {720111}} (\bibinfo {year} {2009})}\BibitemShut
  {NoStop}%
\bibitem [{\citenamefont {Bonaccorso}\ \emph {et~al.}(2010)\citenamefont
  {Bonaccorso}, \citenamefont {Sun}, \citenamefont {Hasan},\ and\ \citenamefont
  {Ferrari}}]{Bonaccorso_NatPhot10}%
  \BibitemOpen
  \bibfield  {author} {\bibinfo {author} {\bibfnamefont {F.}~\bibnamefont
  {Bonaccorso}}, \bibinfo {author} {\bibfnamefont {Z.}~\bibnamefont {Sun}},
  \bibinfo {author} {\bibfnamefont {T.}~\bibnamefont {Hasan}}, \ and\ \bibinfo
  {author} {\bibfnamefont {A.}~\bibnamefont {Ferrari}},\ }\href@noop {}
  {\bibfield  {journal} {\bibinfo  {journal} {Nat. Photonics}\ }\textbf
  {\bibinfo {volume} {4}},\ \bibinfo {pages} {611} (\bibinfo {year}
  {2010})}\BibitemShut {NoStop}%
\bibitem [{\citenamefont {Milonni}\ and\ \citenamefont
  {Eberly}(1988)}]{Eberly_Lasers88}%
  \BibitemOpen
  \bibfield  {author} {\bibinfo {author} {\bibfnamefont {P.~W.}\ \bibnamefont
  {Milonni}}\ and\ \bibinfo {author} {\bibfnamefont {J.~H.}\ \bibnamefont
  {Eberly}},\ }\href@noop {} {\emph {\bibinfo {title} {Lasers}}},\ \bibinfo
  {edition} {1st}\ ed.\ (\bibinfo  {publisher} {Wiley-Interscience},\ \bibinfo
  {address} {New York},\ \bibinfo {year} {1988})\BibitemShut {NoStop}%
\end{thebibliography}

%

\end{document}